\documentclass[12pt]{iopart}
\usepackage{graphicx}  
\usepackage{subfigure}
\begin{document}
\jl{1}
\title[Thresholds in layered neural networks]
        {Thresholds in layered neural networks with variable activity}
\author{D Boll\'{e}\dag\footnote[3]{Also at Interdisciplinair 
Centrum voor Neurale Netwerken, K.U.Leuven, Belgium\\e-mail:
desire.bolle@fys.kuleuven.ac.be, massolo@logikos.it} and 
G Massolo\dag\ddag}
\address{\dag Instituut voor Theoretische Fysica,
K.U. Leuven, B-3001 Leuven, Belgium}
\address{\ddag Universit\`a  degli Studi dell'Insubria,
Facolt\`a di Scienze, I-22100 Como, Italy}

\begin{abstract}
The inclusion of a threshold in the dynamics of layered neural networks
with variable activity is studied at arbitrary temperature. In
particular, the effects on the retrieval quality of a self-controlled
threshold obtained by forcing the neural activity to stay equal to the
activity of the stored patterns during the whole retrieval process, are
compared with those of a threshold chosen externally for every loading
and every temperature through optimization of the mutual information
content of the network. Numerical results, mostly concerning low activity
networks are discussed.
\end{abstract}

\pacs{64.60Cn, 87.10+e, 02.50-r}

\section{Introduction}
Recently, the introduction of a threshold in the dynamics of neural
networks with low activity is discussed again by several authors
\cite{G,KA,DB} (and references therein). Diluted models \cite{G,DB} and 
models for sequential patterns \cite{KA} have been looked at. In all
cases it is found that the retrieval quality -- overlap, basin of
attraction, critical storage capacity, information content -- depends on
the methods of activity control employed.
New insights in the dynamical properties of these models have been
obtained and new suggestions have been put forward for the choice of 
threshold functions in order to get enhanced retrieval.      
In this context it is interesting to study low activity (or in other
words sparsely coded) layered neural networks because as is common
knowledge by now, exactly these models are used in many applications in
several areas of research. 

Sparsely coded models have a very large storage capacity behaving as
$1/(a\ln a)$ in the limit $a$ going to $0$, where $a$ is the pattern
activity (see, e.g., \cite{W,P,Ga,Ok} and references therein). However,
for low activity the basins of attraction might become very small and
the information content in a single pattern is reduced. For the models
mentioned above these drawbacks
can be avoided and an optimal retrieval performance can be reached by   
introducing an appropriate threshold in the dynamics \cite{G,KA,DB,Ok}. 

In the layered models discussed in the sequel we follow two different
approaches.  The first one consists
in forcing the neural activity to be the same as the activity of the
stored patterns during the whole retrieval process. In order to
guarantee this we introduce a time-dependent threshold in the dynamics  
chosen as a function of the noise and the pattern activity in the
network and adapting itself autonomously in the course of the time
evolution. This is the self-control method proposed in \cite{DB}. 

The second approach chooses a threshold by optimising the information
content of the network since for very small pattern activities the
number of active neurons and the information represented by a single
pattern decreases. The relevant quantity we use here is the mutual
information function \cite{DB,Sh,Bl} and the threshold will be called
the optimal threshold. Here the threshold is time-independent and
externally chosen for every loading and every temperature.
Both methods are compared for zero and non-zero temperatures for
networks with various activities. 
 
The rest of this paper is organised as follows. In Section 2 we
introduce the layered network model and define the relevant order
parameters. Section 3 presents the dynamical evolution equations for
these order parameters obtained by the probabilistic signal-to-noise
ratio analysis. In Section 4 we discuss the different threshold mostly
in the 
context of low activity. In Section 5 we present numerical results at
zero and non-zero temperatures. Finally we end with some concluding
remarks in Section 6. 


\section{The model}
Consider a neural network composed of binary neurons arranged in layers, each
layer containing $N$ neurons. A neuron can take values ~$\sigma_{i}(t) \in
\{0,1\}$ where $t=1,\ldots,L$ is the layer index and~ $i=1, \ldots ,N$~
labels the site. Each neuron in layer $t$ is unidirectionally
connected to all neurons on layer $t+1$.
We want to store $p=\alpha N$ patterns  $\{\xi_i^\mu(t)\},
{{i=1,\ldots,N}, ~{\mu=1,\ldots,p}}$ on each layer $t$, taking the
values $\{0,1\}$. They are assumed to be
independent identically distributed random variables (i.i.d.r.v.) with respect
to $i$, $\mu$ and $t$, determined by the probability distribution:
$p(\xi_i^\mu (t))=a\delta(\xi_i^\mu (t)-1)+(1-a)\delta(\xi_i^\mu (t))$.
From this form we find that the
expectation value and the variance of the patterns are given by 
   $
     E[\xi_i^\mu (t)]=E[\xi_i^\mu (t)^2]=a~.
   $
Moreover, no statistical correlations occur, in fact for $\mu\neq\nu$
the covariance vanishes: 
    $
     \mbox{\rm Cov}(\xi_i^\mu (t),\xi_i^\nu (t)) \equiv
     E[\xi_i^\mu (t)\xi_i^\nu (t)]-E[\xi_i^\mu (t)] E[\xi_i^\nu(t)]=0~.
    $
In the sequel it will be convenient to make the change of variables~ 
${\eta_i^\mu(t)}=\xi_i^\mu (t)-a$ ~such that the interesting expectation
values are
 $   E[{\eta_i^\mu(t)}]=0$ and 
 $   E[{\eta_i^\mu(t)}^2]=a(a-1)\equiv\tilde{a}~. $

The state $\sigma_{i}(t+1)$ of neuron $i$ on layer $t+1$ is determined
by the state of the neurons on the previous layer $t$ according to the  
stochastic rule
\begin{equation}  
       \label{eq:stoc}
 \fl \hspace{1cm} P(\sigma_{i}(t+1)\mid \sigma_{1}(t), \ldots ,\sigma_{N}(t))
             = \{1+\exp[2(2\sigma_i(t+1)-1) \beta{h_i(t)}]\}^{-1}.
\end{equation}
The parameter $\beta =1/T$ controls the stochasticity of the network
dynamics, it measures the noise level.
Given the configuration  $\{\sigma_i(t)\};{i=1,\ldots,N}$ on layer $t$,
the local field ${h_i(t)}$ in site $i$ on the next layer $t+1$ is given by 
\begin{equation} 
       \label{eq:h}
   h_i(t)= \sum_{j=1}^{N} 
          J_{ij}(t)(\sigma_i(t) -a)-\theta(t) 
\end{equation} 
with $\theta(t)$ the threshold to be specified later. 
The couplings $J_{ij}(t)$ are the synaptic strengths of the interaction  
between neuron $j$ on layer $t$ and neuron $i$ on layer $t+1$. They
depend on the stored patterns at different layers according to the
covariance rule 
\begin{equation}   
       \label{eq:j}
   J_{ij}(t)=\frac{1}{N\tilde{a}} \sum_{j=1}^{N} 
             (\xi_i^\mu (t+1)-a)(\xi_j^\mu (t)-a)~.
\end{equation}
These couplings then permit to store sets of patterns to be retrieved by
the layered network.
We remark that in the limit $T\rightarrow 0$ the updating rule
(\ref{eq:stoc}) reduces to the deterministic form
\begin{equation}  
        \label{eq:step}
    \sigma_i(t+1)=\Theta({h_i(t)})
\end{equation}
where $\Theta(x)$ is the standard step function taking the value $\{0,1\}$.

We take parallel updating. The dynamics of this network is defined as
follows (see \cite{DKM,BSV}  and references therein).
Initially the first layer (the input) is externally set in some fixed
state. In response to that, all neurons of the second layer update
synchronously at the next time step, according to the stochastic rule
(\ref{eq:stoc}), and so on.
Layered feed-forward networks allow an exact analytic treatment of 
their parallel dynamics stemming from the independent choice of the     
representations of the patterns on different layers.
By exact analytic treatment we mean that, given the configuration of the
first layer as initial state, the configuration on layer $t$ that
results from the dynamics is predicted by recursion formulas for the
relevant order parameters. This configuration is known through the calculation
of macroscopic quantities obtained by averaging over the thermal noise
associated with the dynamics, as well as over the random choice of the stored 
patterns.

The relevant order parameters measuring the quality of retrieval are the 
{\it main overlap} of the microscopic state of the network and the $\mu$-th
pattern, and the {\it neural activity} of the neurons   
\begin{equation}
    \label{M(t)}
  M_N^\mu(t)  =
    \frac{1}{N\tilde{a}}\sum_{i=1}^N{\eta_i^\mu(t)}
                     (\sigma_i(t) -a),
    \qquad
   q_N(t) =
      \frac{1}{N}\sum_{i=1}^N \sigma_i(t)~.
\end{equation} 
These order parameters determine the Hamming distance between the state
of the network and the pattern $\{\xi_i^\mu(t)\}$
\begin{equation}
     \label{Ham}
   d_H(\xi^\mu(t),\sigma(t))=\frac{1}{N}\sum_{i=1}^N
            \left[\xi_i^\mu (t)-\sigma_i(t)\right]^2~.
\end{equation}
It is known that the Hamming distance is a good measure for the
performance of a network when the neural activity $a \sim 1/2$. For low
activity networks, however, it does not give a complete description of
the information content \cite{DB}. Therefore, the mutual information
function $I(\sigma_i(t);\xi_i^\mu (t))$ has been introduced  \cite{DB,Bl} 
\begin{equation}  
    \label{eq:inf}
   I(\sigma_i(t);\xi_i^\mu (t))=
       S(\sigma_i(t))-\langle S(\sigma_i(t)|\xi_i^\mu (t))\rangle
                                 _{\xi^{\mu}(t)}
\end{equation}
where $\xi_i^{\mu}(t)$ is considered as the input and $\sigma_i(t)$ as
the output with $S(\sigma_i(t))$ its entropy and
$S(\sigma_i(t)|\xi_i^\mu (t))$ its conditional entropy, viz.
\begin{eqnarray} 
      \label{eq:en}
  S(\sigma_i(t))=-\sum_\sigma p(\sigma_i(t))\ln[p(\sigma_i(t))]\\ 
      \label{eq:enc}
  S(\sigma_i(t)|\xi_i^\mu (t))=
       -\sum_\sigma p(\sigma_i(t)|\xi_i^\mu (t))
                \ln[p(\sigma_i(t)|\xi_i^\mu (t))]~.
\end{eqnarray}
Here $p(\sigma_i(t))$ denotes the probability distribution for the
neurons at time $t$ and $p(\sigma_i(t)|\xi_i^\mu (t))$ indicates the
conditional probability that the $i$-th neuron is in a state
$\sigma_i(t)$ at time $t$ given that the $i$-th site of the stored
pattern to be  retrieved is $\xi_i^\mu (t)$.


\section{Dynamics at arbitrary temperature}
We suppose that the initial configuration $\{\sigma_i(1)\}$ 
is a collection of i.i.d.r.v. with average and variance given by 
$
      E[\sigma_i(1)]=E[(\sigma_i(1))^2]=q_0~.
$
We furthermore assume that this configuration is correlated with only
one stored pattern, say pattern $\mu=1$, such that
\begin{equation}
    \label{eq:cov}
   \mbox{\rm Cov}(\xi_i^\mu (1),\sigma_i(1))=\delta_{\mu,1}~M_0^1~\tilde{a}~.
\end{equation}
We then obtain the order parameters (\ref{M(t)}) at the initial time
step $t=1$ in the thermodynamic limit by the law of large numbers (LLN).
For the main overlap we have
\begin{equation}
 \fl    M^\mu(1)\equiv\lim_{N\rightarrow\infty}M_N^\mu(1)
             \stackrel{LLN}{=}
        \frac{1}{\tilde{a}} E[{\eta_i^\mu(1)}(\sigma_i(1)-a)]= \
             \frac{1}{\tilde{a}}\mbox{\rm Cov}(\xi_i^\mu (1),\sigma_i(1))
              =\delta_{\mu,1}M_0^1
	      \label{eq:M}
\end{equation}
and for the neural activity 
\begin{equation}    
      \label{eq:q}
    q(1)\equiv\lim_{N\rightarrow\infty} q_N(1)
          \stackrel{LLN}{=}E[\sigma_i(1)]=q_0~.
\end{equation}

The evolution equations governing the dynamics are then obtained
following the methods based upon a {\it signal-to-noise} analysis of the
local field (see, e.g., \cite{DKM}-\cite{MD3} for the case
without threshold and without bias, i.e., $a=1/2$). The local field
is split as the sum of a signal (from the condensed pattern $\mu=1$) and
a noise (from the non-condensed patterns $\mu >1$). For a recent
overview comparing various architectures we refer to \cite{BJSP}. Since
the method is standard by now we only write down the final results. At
zero temperature we obtain for a general time step 
\begin{eqnarray}
\fl  \hspace{0.5cm}   M^1(t+1)=1-\frac{1}{2}\left\{\mbox{\rm erfc}
       \left(\frac{(1-a)M^1(t)-\theta(t)}{\sqrt{2\alpha D(t)}}\right)+
                    \mbox{\rm erfc}
       \left(\frac{aM^1(t)+\theta(t)}{\sqrt{2\alpha D(t)}}\right)\right\}
     \label{eq:mrec}    \\
\fl  \hspace{0.5cm}   q(t+1)=aM^1(t+1)
          + \!\frac{1}{2}\mbox{\rm erfc}\!
	 \left(\frac{aM^1(t)+\theta(t)}{\sqrt{2\alpha D(t)}}\!\right)
		       	\label{eq:qrec}	       \\  
\fl   \hspace{0.5cm}  D(t+1)=Q(t+1)+\frac{1}{2\pi\alpha}
             \left\{a\exp\left(-\frac{(\theta(t)-(1-a)M^1)^2}
	                {2D(t)\alpha}\right) \right. 
			\nonumber \\
                + \left. (1-a)\exp\left(-\frac{(\theta(t)+aM^1)^2}
                  {2D(t)\alpha}\right)\right\}^2
		  \label{eq:drec}
 \end{eqnarray}
where $Q(t)=(1-2a)q(t)+a^2$ and $D(t)$ is the variance of the residual
overlap containing the influence of the non-condensed patterns $\mu>1$.
The residual overlap is defined as
\begin{equation} 
    \label{eq:rn}
  r_N^\mu(t)=\frac{1}{\sqrt{N\tilde{a}}} 
                \sum_{i=1}^N{\eta_i^\mu(t)}(\sigma_i(t)-a)\,, \quad
		\mu>1,
\end{equation}
and causes the intrinsic noise in the dynamics of the main overlap
$M^1(t)$.
Finally, $\mbox{\rm erf}(x)=(2/\sqrt{\pi}) \int_0^x dy \exp(-y^2)$.

For non-zero temperatures thermal averages denoted by $\langle \cdots
\rangle$ have to be taken in agreement
with the distribution (\ref{eq:stoc}) such that
\begin{equation}
   \label{eq:sav}
 \langle \sigma_i(t+1)\rangle
 =\frac{1}{2}\left[1+\tanh(\beta\langle{h_i(t)}\rangle)\right]
\end{equation}
and
\begin{equation}
\label{eq:Mav}
   M^\mu(t)\equiv\lim_{N\rightarrow\infty} \langle M_N^\mu(t) \rangle~,
   \quad
   q(t)\equiv\lim_{N\rightarrow\infty} \langle q_N(t) \rangle~.
\end{equation}
The stochastic dynamics can then be described through the following
equations for the order parameters 
\begin{eqnarray} 
\fl \hspace{0.5cm}    M^1(t+1) =  \frac{1}{2}\left\{\int{\cal D} x
       \tanh\left[\beta((1-a)M^1(t)-\theta(t)+\sqrt{\alpha D(t)}\,x)\right]
             \right. \nonumber \\  
	     + \left.  \int{\cal D} x
       \tanh\left[\beta(-aM^1(t)-\theta(t)+ \sqrt{\alpha D(t)}\,x)\right]
                   \right\} 
              \label{eq:a}\\
\fl \hspace{0.5cm}     q(t+1)= aM^1(t+1)+\frac{1}{2} 
           \left\{ 1+\int{\cal D} x
     \tanh\left[\beta(-aM^1(t)-\theta(t)+\sqrt{\alpha D(t)}\,x)\right]
                        \right\}
         \label{eq:b} \\
\fl \hspace{0.5cm}      D(t+1)= Q(t+1) + 
        \frac{\beta}{2}\left\{1-a\int{\cal D} x\tanh^2\beta\left[(1-a)M^1(t)
            -\theta(t)+\sqrt{\alpha D(t)}\,x\right] \right. 
	    \nonumber \\
	   - \left. (1-a)\int{\cal D} x\tanh^2\beta
	\left[-aM^1(t)-\theta(t)+\sqrt{\alpha D(t)}\,x\right]\right\}
\end{eqnarray}  
where ${\cal D} x$ is the Gaussian measure ${\cal D} x= dx
(2\pi)^{-1/2}\exp(-x^2/2)$. 


\section{Thresholds}
\subsection{Low activity and self-control} \label{sec:sc}
In the limit of low activity it has been emphasized already in the study
of extremely diluted and fully connected architectures that one
should try to keep the pattern activity of the network during the
retrieval process the same as the one for the memorized patterns
\cite{G,DB,Ok,AGS,A,BDS,HU}. 
Also for the layered model considered here one easily finds for 
fixed $\alpha$ and zero threshold by using 
eqs.~(\ref{eq:mrec})-(\ref{eq:qrec})
that in the limit $a \rightarrow 0$ the neural activity behaves as     
$q(t)\sim\frac{1}{2}+aM^1(t)$ and always tends to $1/2$. The way to avoid
 this is to
choose, given $a$, the capacity $\alpha$ such that ${aM^1(t)}\sim  
\sqrt{2\alpha D(t)}$ but this means that when
$a$ decreases the critical capacity $\alpha_c$ is going to decrease too.
In fact, numerical experiments on the layered model show
that for $\theta=0 $ and $a\approx 10^{-3}$, $\alpha_c \approx 10^{-4}$. 
Similar considerations stay valid for non-zero temperatures.

Therefore, in the retrieval process we need to control the neural activity 
and keep it, at each layer, the same as the one 
for the stored patterns: $q(t)=a$. For a network with low activity this
requires the introduction of a threshold $\theta(t)$ in the definition of the 
local field (\ref{eq:h}).
For the extremely diluted model a time-dependent threshold has been chosen 
in \cite{DB} as a function of the noise in the system and the pattern
activity, adapting itself in the course of the time evolution. The novel
idea was to let the network itself autonomously counter the residual
noise at each time step of the dynamics without having to impose any 
external constraints.

In the following we start from an analogous general form for this
self-control threshold
\begin{equation}   
    \label{eq:thr}
     \theta(t)_{sc}=c(a)\sqrt{\alpha D(t)}
\end{equation}
where we recall that $D(t)$ is the variance of the noise contribution in the
local field. For the determination of $c(a)$ we consider the form 
(\ref{eq:qrec}) for the layered architecture and require that the term 
\begin{equation}
  \mbox{\rm erfc}\left(\frac{aM^1(t)+\theta(t)}{\sqrt{2\alpha D(t)}}\right)
     = \mbox{\rm erfc}\left(\frac{aM^1(t)}{\sqrt{2\alpha D(t)}}+
          \frac{c(a)}{\sqrt{2}}\right)
         \sim  \frac{e^{-c(a)^2/2}}{c(a)\sqrt{2\pi}}
\end{equation}	
must vanish faster than $a$. This can be realised by choosing 
$c(a)=\sqrt{-2\ln a}$. 
We furthermore remark that in the low activity limit the recursion relation 
(\ref{eq:drec}) for $D(t+1)$ leads to $D(t+1)\sim Q(t+1)$.
This shows explicitly that in this limit the result for
the layered model is similar to the one for the extremely diluted 
model \cite{DB}. Indeed, we intuitively expect that in the limit of low
activity all models roughly behave in the same way.

The line of arguments above is also valid at arbitrary temperature. In the
limit of low activity it is straightforward to show that the second
term on the r.h.s. of eq.~(\ref{eq:b})
vanishes faster than the activity  $a$.

We recall that this self-control threshold (\ref{eq:thr}) is a macroscopic 
parameter, thus
no average must be taken over the microscopic random variables at each
time step $t$. We have in fact a mapping with a threshold changing each
time step, but no statistical history intervenes in this process.

In the next section we study explicitly the influence of this
threshold on the retrieval quality of the network dynamics. For the
extremely diluted model \cite{G,DB} and in the case of sparsely coded
sequential patterns \cite{KA} it has been shown that this retrieval
quality is considerably improved for low activity. In the case of the  
extremely diluted model this improvement also works for not so low
activity \cite{DB}. Furthermore, although the form of the threshold has
been derived at zero temperature, we also want to find out whether 
it works at finite temperatures.


\subsection{Optimising the mutual information}
We have argued that the mutual information function (\ref{eq:inf})
is a better concept than the Hamming distance in order to measure the
retrieval 
quality especially in the limit of low activity. So, a second type of  
threshold we introduce is obtained by optimising this mutual information.

We start by calculating the mutual information for the case at hand
using the eqs.~(\ref{eq:inf}), (\ref{eq:en}) and (\ref{eq:enc}).
In the sequel we drop the index $t$. Because of the mean-field
character of our model the following formula hold for every site index
$i$  on each 
layer $t$. After some algebra we find for the conditional probability
\begin{equation}
    p(\sigma|\xi)=[\gamma_0\xi+(\gamma_1-\gamma_0)\xi]\delta(\sigma-1)+
       [1-\gamma_0-(\gamma_1-\gamma_0)\xi]\delta(\sigma)
\end{equation}
where $\gamma_0=q-aM^1$ and $\gamma_1=(1-a)M^1+q$, and where the $M^1$
and $q$ 
are precisely the order parameters (\ref{M(t)}) for $N \rightarrow
\infty$.
Using the probability distribution of the patterns we obtain
\begin{equation}
   p(\sigma)=q\delta(\sigma-1)+(1-q)\delta(\sigma)~.
\end{equation}   
Hence the entropy (\ref{eq:en}) and the conditional entropy
(\ref{eq:enc})  become
\begin{eqnarray}
           S(\sigma)=&-&q\ln q -(1-q)\ln(1-q) \\
           S(\sigma|\xi)=&-&[\gamma_0+(\gamma_1-\gamma_0)\xi]
	             \ln[\gamma_0+(\gamma_1-\gamma_0)\xi] 
		     \nonumber \\
                &-&[1-\gamma_0-(\gamma_1-\gamma_0)\xi]
                      \ln[1-\gamma_0-(\gamma_1-\gamma_0)\xi]~.
\end{eqnarray}		      
By averaging the conditional entropy over the pattern $\xi$ we get
\begin{equation}
 \fl   \langle S(\sigma|\xi)\rangle _\xi
      =-a[\gamma_1\ln\gamma_1+(1-\gamma_1)\ln(1-\gamma_1)]
        -(1-a)[\gamma_0\ln\gamma_0+(1-\gamma_0)\ln(1-\gamma_0)]
\end{equation}	    
such that the mutual information function (\ref{eq:inf}) for the layered
model is given by
\begin{eqnarray}
     \label{eq:II}
\fl \hspace{1cm}      I(\sigma;\xi) = -q\ln q -(1-q)\ln(1-q)
          +a[\gamma_1\ln\gamma_1+(1-\gamma_1)\ln(1-\gamma_1)]
	  \nonumber\\
	     \label{eq:Ifin}
	 +(1-a)[\gamma_0\ln\gamma_0+(1-\gamma_0)\ln(1-\gamma_0)]~. 
\end{eqnarray}
At time $t$ the mutual information function depends on the main
overlap $M^1(t)$, the neural activity  $q(t)$, the pattern activity $a$,
the storage capacity  $\alpha$ and the  inverse temperature $\beta$. The
evolution of 
the main  overlap and of the neural activity (eqs.~(\ref{eq:mrec}), 
(\ref{eq:qrec}) for zero temperature and (\ref{eq:a}), (\ref{eq:b}) for
arbitrary temperature) depends on the specific choice of the threshold
in the definition of the local field (\ref{eq:h}). We consider a
time-independent threshold $\theta(t)=\theta$ 
and calculate the value of (\ref{eq:Ifin}) at equilibrium for
fixed $a$, $\alpha$, $M_0$, $q_0$ and $\beta$. The optimal choice for
this threshold
chosen at equilibrium, $\theta=\theta_{opt}$, is then
the one for which the mutual information function is maximal.


\section{Results}
We have studied the retrieval properties for the layered
model with $\theta_{sc}$ and $\theta_{opt}$ by numerically solving the
recursion relations derived in section~3 with an activity ranging from
$a=0.001$ to $a=0.3$ at various inverse temperatures $\beta=3,4,5,10,100,
\infty$. We are interested only in the retrieval solutions with $M^1>0$
(in the sequel we drop the superindex 1)
and carrying a non-zero information $I$. The results for zero
and non-zero temperature have been analysed separately. Our main aim is
to study how self-control introduced for extremely diluted networks
also works for  other models, in casu a layered architecture at zero 
temperature, as claimed
in \cite{DB}, and to check whether such a threshold can still be useful at
non-zero temperatures. Moreover, we compare this self-control
method, which is mainly designed for low activity but also works for
higher activities, with the optimization method. Since the latter works
for all values of the activity although it has to be found externally
for every loading and every  temperature. 


\subsection{Zero temperature}
\begin{figure}[t]
   \vspace{1.5cm}
       \centering
      \includegraphics[height=0.5\textwidth,
                       angle=-90]{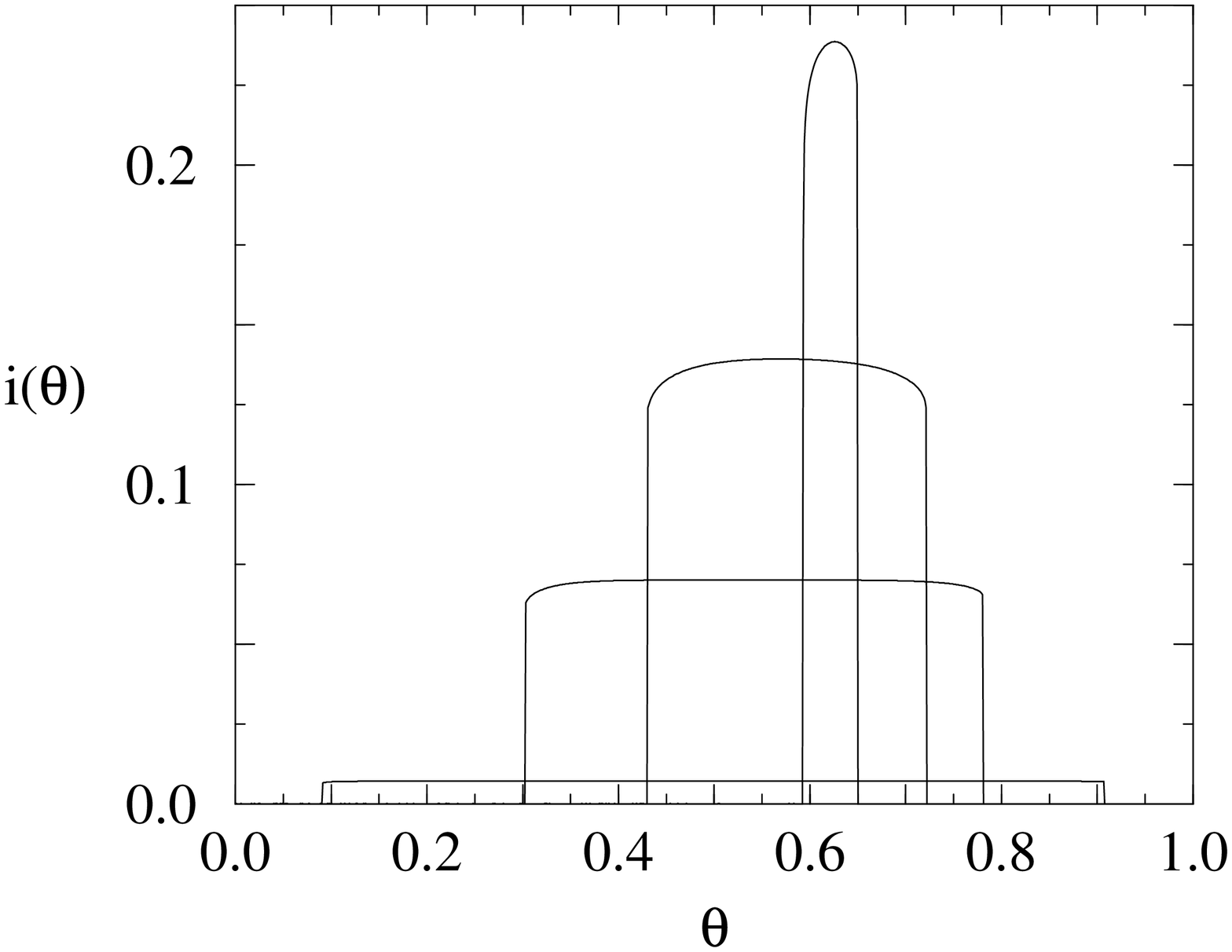}
       \caption{The information $i$ as a function of $\theta$ for $a=0.01$ 
        and several values of the storage capacity $\alpha=0.1,~1,~2,~4$
	(bottom to top).}
       \label{fig:infA}
\end{figure}

In fig.\ref{fig:infA} we have plotted the information content 
$i=\alpha I$ as a function of $\theta$  without self-control or a priori
optimization for pattern activity $a=0.01$ and
different values of the storage capacity $\alpha$. For every value of 
$\alpha$, below its critical value, there is a range for the threshold 
where the information content is different from zero. For any   
choice of the threshold in this range retrieval is possible. This
retrieval range becomes very small when the capacity approaches its
critical value $\alpha_c=4.72$. 

Defining the basin of attraction as the range of initial values $M_0
\in [0,1]$ which lead to the retrieval attractor $M(t)\sim 1$, we remark at
this point that the size of this basin strongly depends on the specific 
choice of the threshold in the retrieval range. Technically it turns out
that the value to be chosen for the latter in order to have the  
largest basin is the minimal $\theta$ in the retrieval range. This, of
course, has to be repeated for every $\alpha$. This threshold 
optimises the information content and is called, as specified 
before,  $\theta_{opt}$.

\begin{figure}
       \centering   
       \subfigure[]{
             \label{fig:1b}
             \includegraphics[height=0.45\textwidth,
                               angle=-90]{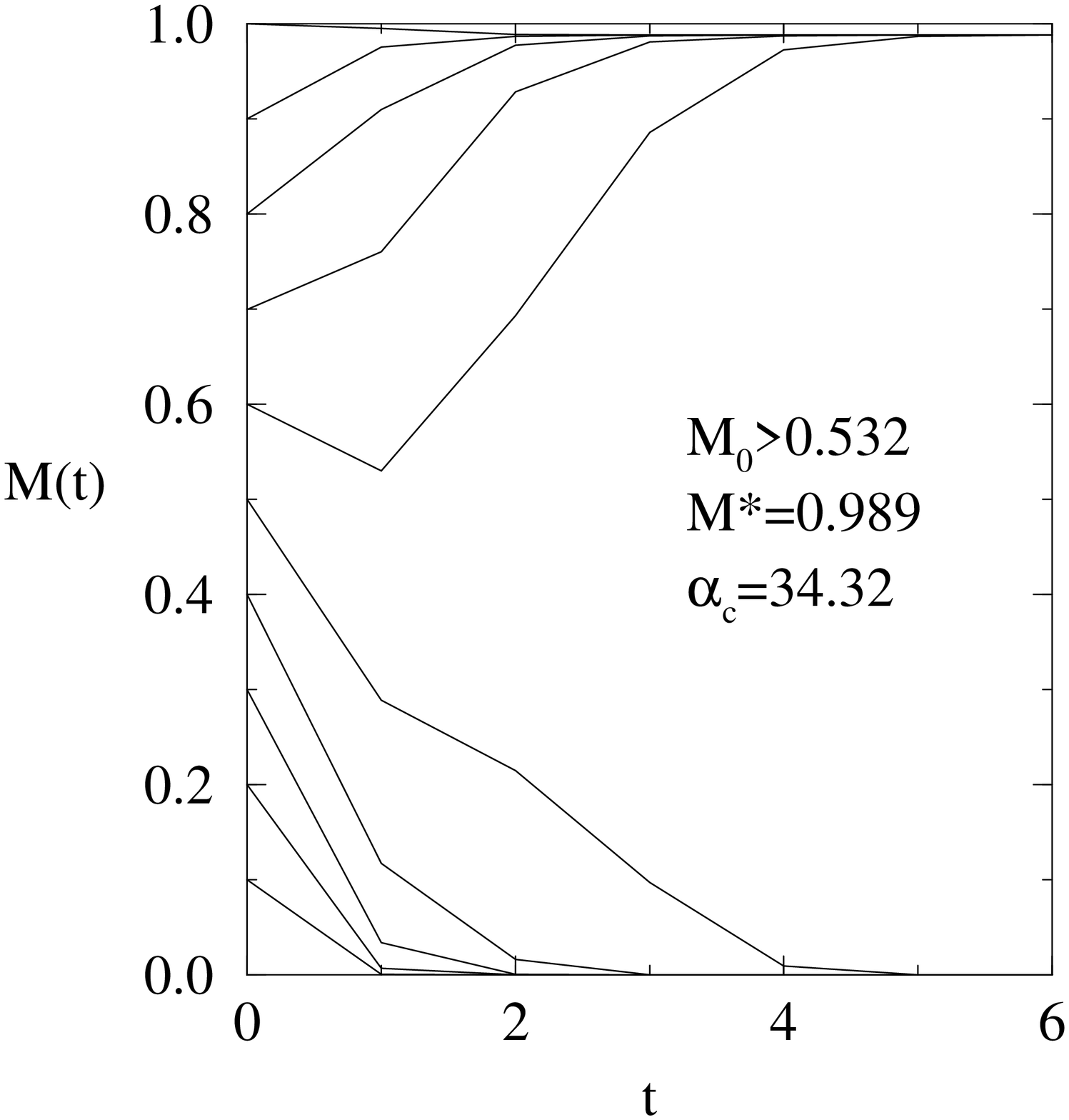}}
     	 \hspace{0.3cm}
	\subfigure[]{
	\label{fig:1c}
             \includegraphics[height=0.45\textwidth,
                              angle=-90]{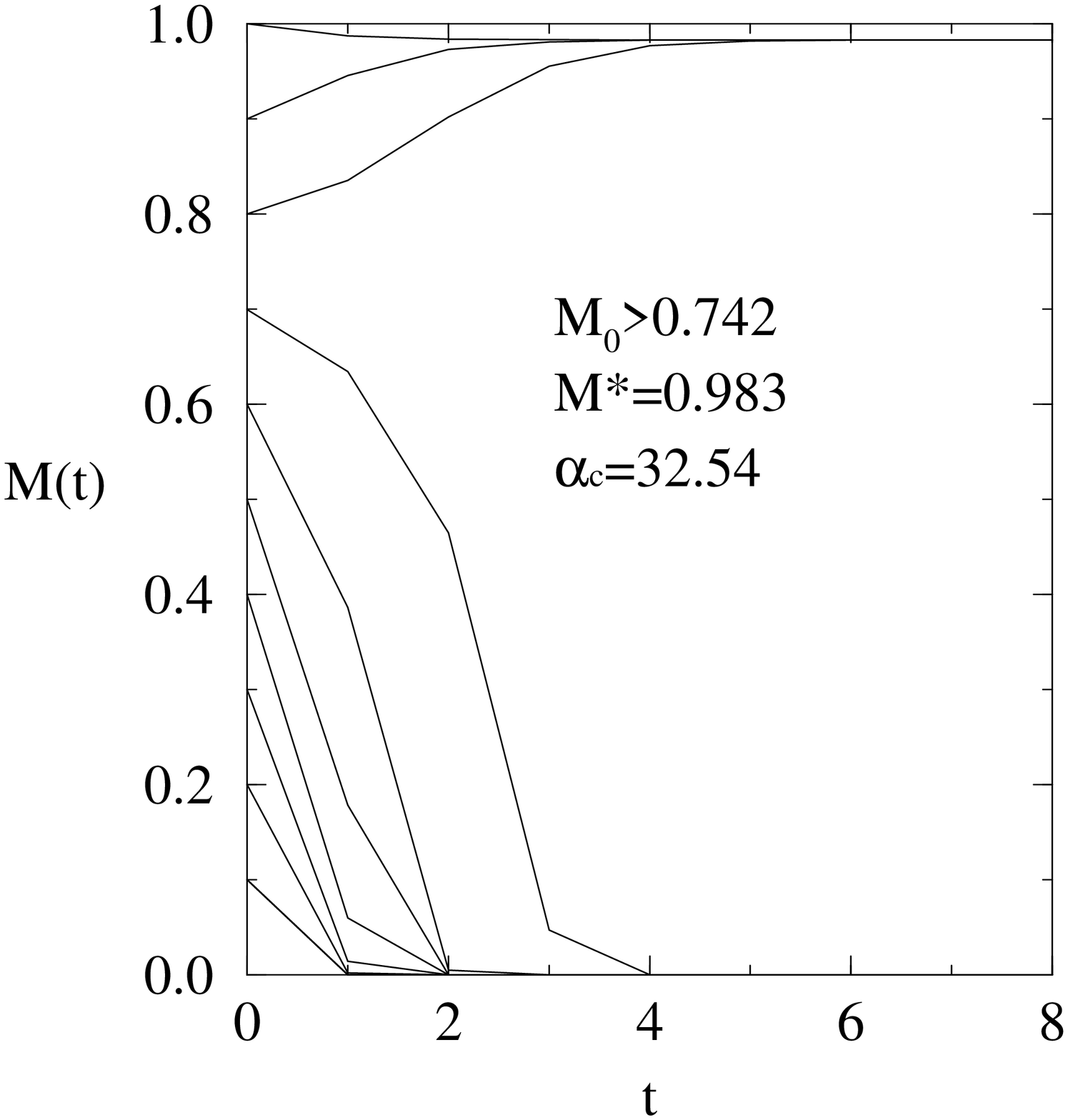}}     
	     \caption{The evolution of the main overlap $M(t)$ for
	     several initial values $M_0$ with $q_0=a=0.001,~\alpha=25$
	     for the self-control model (a) and the optimal threshold
	     model (b).} 
       \label{fig:bas}
\end{figure}

Figure \ref{fig:bas} represents the dynamical evolution of the network. The
retrieval overlap $M(t)$ is shown as a function of time for different initial
values $M_0$, $q_0=0.001=a$ and $\alpha=25$. A self-control threshold 
$\theta_{sc}= [-2 (\ln a) \alpha Q(t)]^{-1/2}$ 
(fig.~\ref{fig:1b}) is compared with an optimal threshold
$\theta_{opt}$
(fig.~\ref{fig:1c}) concerning the values of the minimal $M_0$ for
retrieval, the fixed-point $M^*$  and the critical capacity $\alpha_c$.
It is seen that self-control works better than optimization and both much
better than a zero threshold (where there is no retrieval at all since
$\alpha_c =5.3 \times 10^{-5}$ only). 
This can be interpreted as a result of the property of adaptivity in the
course of the time evolution inherent in the self-control method.

\begin{figure}
    \vspace{1.5cm}
       \centering
             \includegraphics[height=0.5\textwidth,
                              angle=-90]{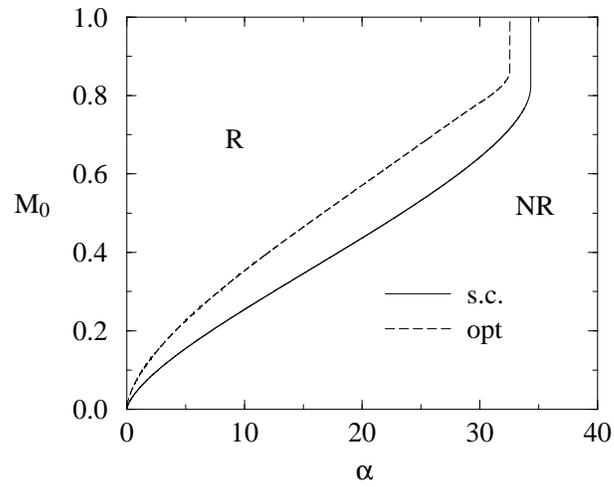}      
	     \caption{The basin of attraction as a function of $\alpha$
	       for $a=0.001$ for the self-control model (full line) and
	       the optimal threshold model (dashed line).} 
      \label{fig:3}
\end{figure}

In fig.\ref{fig:3}  the retrieval phase diagram is illustrated for
$a=0.001$ and $q_0=a$. 
In the low activity limit the basin of attraction is substantially improved
by self-control even near the border of the critical storage. Hence, also the
storage capacity is larger with self-control. Furthermore, we have compared 
these curves with the one for a model without threshold  in the low
activity limit. Since we find a very small storage capacity (of order
$~10^{-4}$) such a network without threshold has very little interest.
    
\begin{figure}[t]
       \centering
       \includegraphics[height=0.7\textwidth,
                      angle=-90]{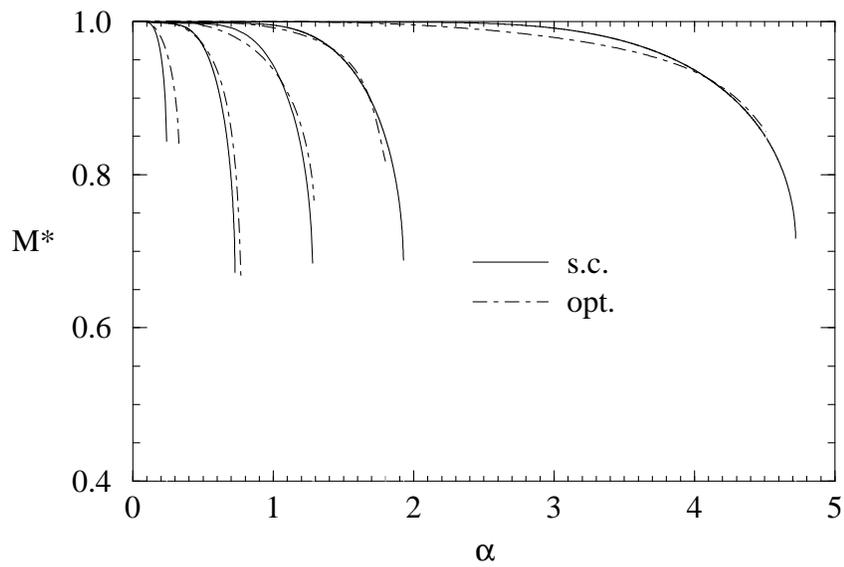}
       \caption{The retrieval fixed-points $M^*$ as a function of $\alpha$
         for the self-control model (full line) and the  optimal threshold 
	 model (dashed line) with  decreasing  pattern activity: 
	      $a=0.3, 0.1, 0.05, 0.03, 0.01$ (from left to right).}
       \label{fig:4}
\end{figure}

Plotting the retrieval fixed-point  $M^*$ as a function of
$\alpha$ we have found a first order transition from the retrieval phase 
($M^*>0$) to the non-retrieval one ($M^*=0$), see fig.~\ref{fig:4} for
different values of $a$.
We remark that the curves for $a=0.001$ are out of the scale of this figure.
In that case  we  find  $\alpha_c=34.32$ and $M^*\sim1$ for $0<\alpha<20$. 
We  compare the fixed-point behaviour found with self-control 
(solid lines) with  the results obtained by choosing the threshold
through the optimization of the mutual information function (dashed lines). 
Roughly speaking, self-control is the best choice for activities 
below  $0.05$. For $a$ above this value, but still small compared with a
homogeneous distribution $a=1/2$, e.g. $a=0.3$, 
self-control continues to perform quite well, however it ceases to be better 
than optimization. 
 
Finally, we have studied the leading behaviour of the critical capacity
in the limit $a \rightarrow 0$. We have found that 
$\alpha_c(a)\sim (a|\ln{a}|)^{-1}$. This is consistent with former
studies on other low activity models (see \cite{G,DB} and references
therein). Moreover, we remark that for $a$ in the range 
$(10^{-4},10^{-3})$  the proportionality coefficient seems to be
constant and given by $0.25$.


\subsection{Non-zero temperature}   
Since self-control is completely autonomous and since it improves the
retrieval quality also for not so sparse networks it is worth checking 
how it performs for non-zero temperatures. Also in this case we compare
it with the optimal threshold for which we recall that it has to be
calculated by hand when the network has reached equilibrium for every
loading $\alpha$ and every inverse temperature $\beta$.

\begin{figure}
   \vspace{1cm}
        \centering
        \subfigure[$\beta=3,~ 4,~ 5,~ 100$]{
             \label{tesigrafT1}
             \includegraphics[height=0.7\textwidth,
                             angle=-90]{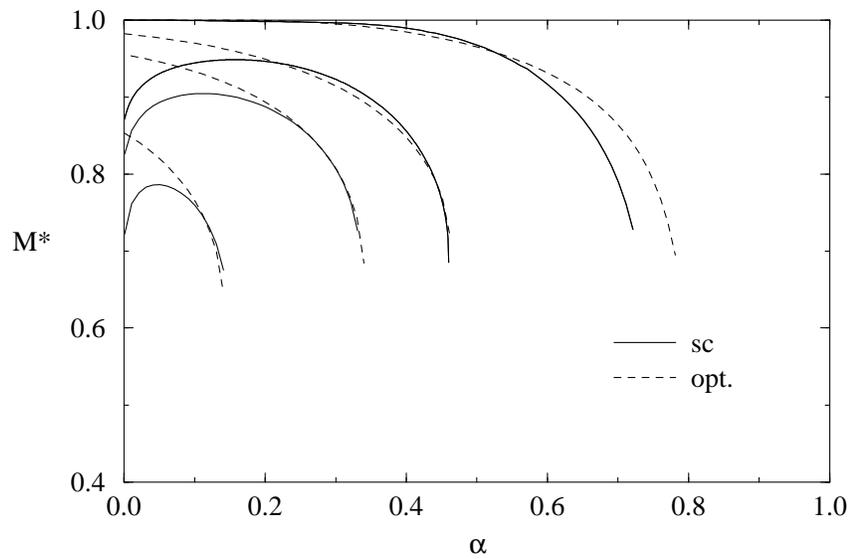}}
             \hspace{10cm}
        \centering
        \subfigure[$\beta=3, ~4,~ 5, ~10,~100$]{
             \label{tesigrafT3}
             \includegraphics[height=0.7\textwidth,
                             angle=-90]{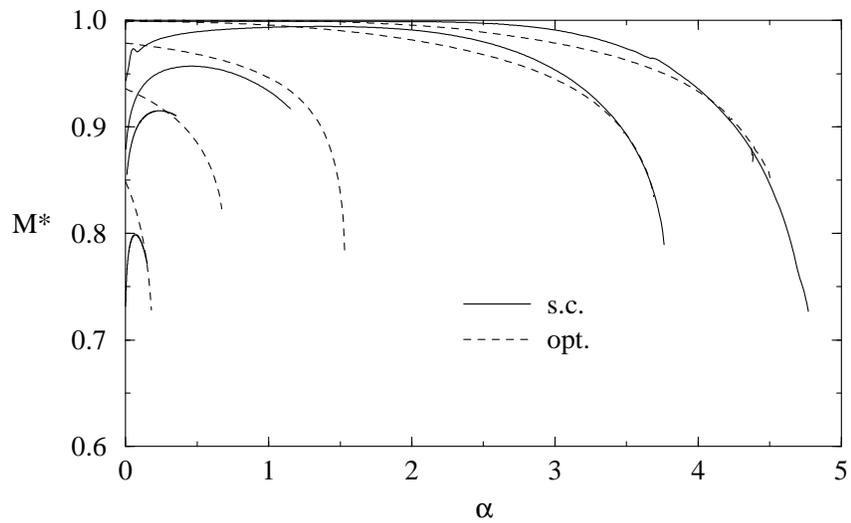}}
             \hspace{10cm}
     \caption{The retrieval fixed-points $M^*$ as a function of
	$\alpha$ for several values of the
	inverse temperature for the self-control model (full line) and
	the optimal threshold model (dashed line) for
		  $a=0.1$ (a) and $a=0.01$ (b).} 
        \label{tesigrafT:14}
\end{figure}

In fig.~\ref{tesigrafT:14} we have studied the retrieval
fixed-points of the main overlap as a function of the storage capacity
for different values of the temperature and of the pattern activity.
The results are plotted for $a=0.1$ (fig.~\ref{tesigrafT1}) and
$a=0.001$ (fig.~\ref{tesigrafT3}) and increasing $\beta$. 
The lines end at the critical capacity where a first-order
transition to the non-retrieval phase occurs.
At this point we recall that also for these non-zero temperatures in
both cases the presence
of a non-zero threshold is strictly necessary in order for the network
to evolve toward the retrieval phase for these storage capacities. 

For $\beta=100$ the results of the deterministic network are found back.
For $a=0.1$ we already know from the previous analysis at zero
temperature that optimization works better than self-control. For
$a=0.01$ the reverse situation is valid.
For smaller $\beta$ and for smaller storage capacities self-control does
 not work as well. Optimization leads to a bigger
value for the retrieval overlap than self-control does. 

For the lowest pattern activity $a=0.01$ (fig.~\ref{tesigrafT3})
self-control works worse for increasing temperature. The critical
capacity of the network with
self-control is smaller than the critical capacity obtained by optimization. 
In fact, for $\beta=3,~4$ it is about  half. 
For  pattern activity below $a=0.01$ the critical capacity with
self-control becomes still smaller and it is smaller than the critical
capacity obtained by optimization.

We can then summarize the peculiar behaviour with self-control 
for small storage capacities as follows. We usually expect
the retrieval fixed-points to have the greatest overlap values at zero storage 
capacity and then to slowly decrease until the critical capacity is 
reached, where there is a phase
transition. This is, indeed, the behaviour at zero temperature with
whatever choice of the threshold. At non-zero temperature this behaviour
is found with the optimisation approach, with the self-control method
the retrieval fixed-points obtain their maximal retrieval overlap not at
zero capacity, but at a higher value.

\begin{figure}[b]
     \vspace{1.5cm}
        \subfigure[]{
             \label{tesigrafT6a}
             \includegraphics[height=0.45\textwidth,
                         angle=-90]{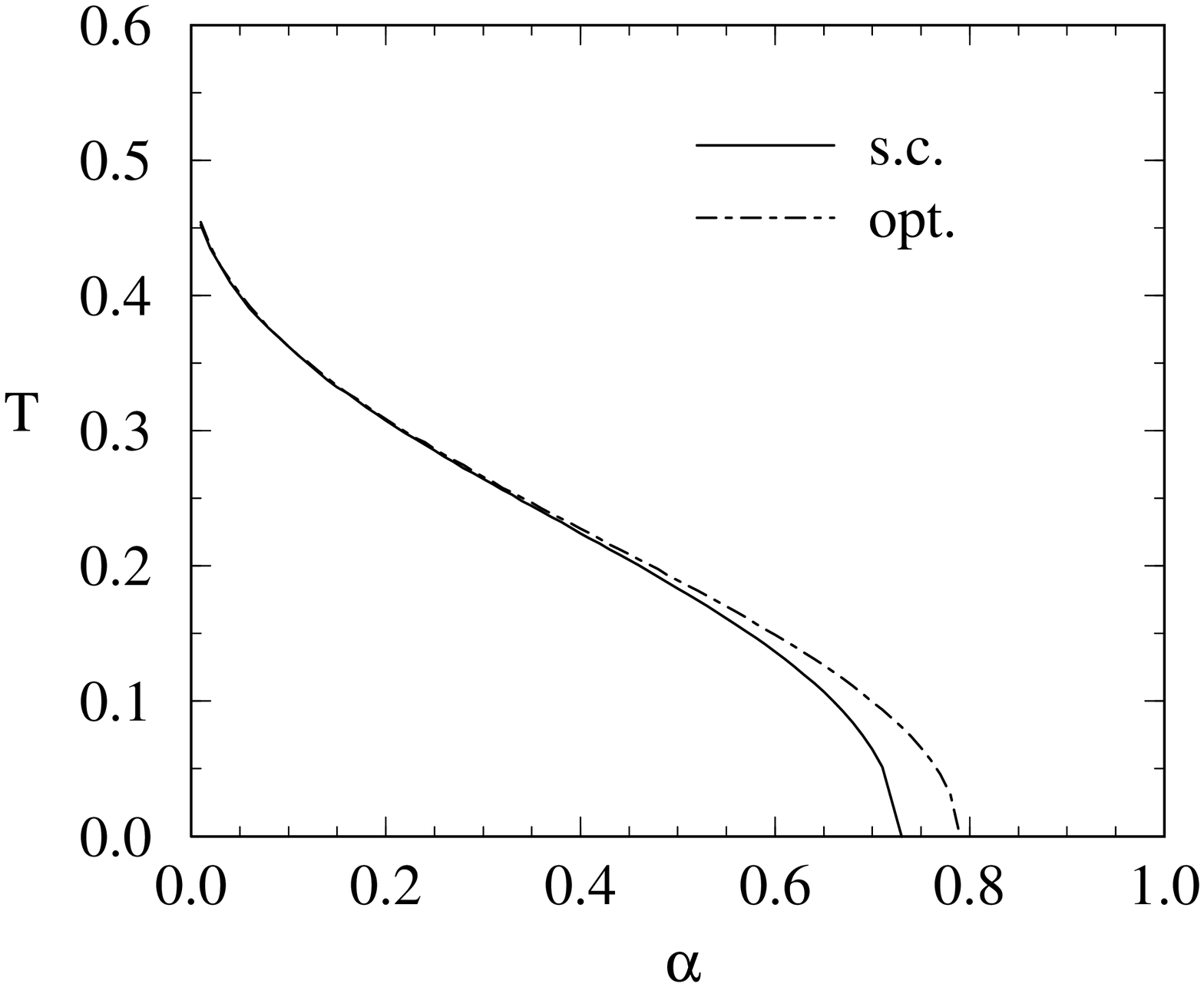}}
        \subfigure[]{
             \label{tesigrafT6b}
             \includegraphics[height=0.45\textwidth,
                         angle=-90]{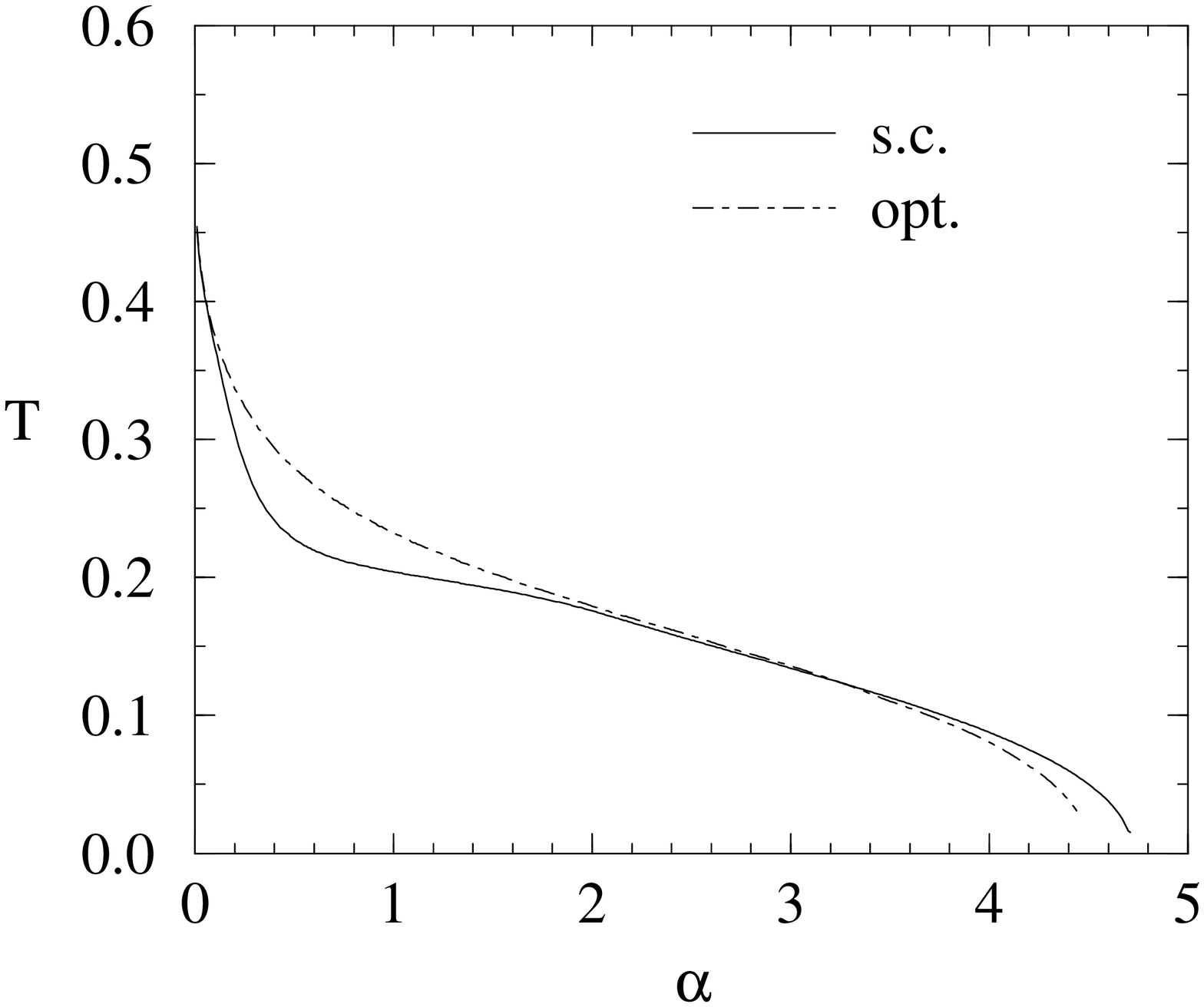}}
	\hspace{0.5cm}
	\centering
        \subfigure[]{
             \label{tesigrafT6d}
             \includegraphics[height=0.45\textwidth,
                         angle=-90]{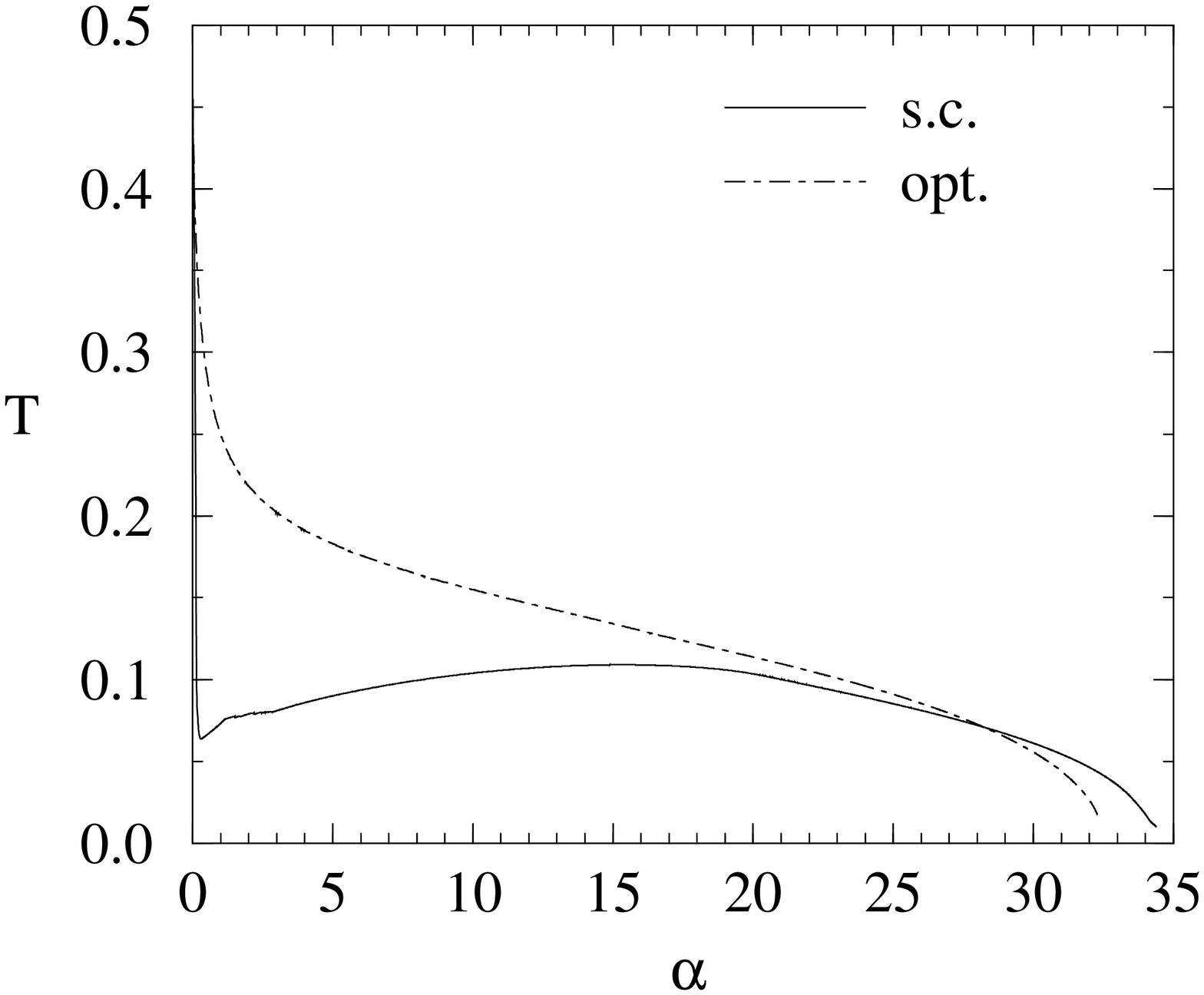}}
        \caption{The temperature-capacity phase diagram for the
	self-control  model (full line) and
	the optimal threshold model (dashed-dotted line) for
		$a=0.1$ (a),  $a=0.01$ (b) and $a=0.001$ (c).}
        \label{tesigrafT6}
\end{figure}

The analysis of the temperature-capacity phase diagram with self-control
and optimization for different values of the pattern activity is
summarized in fig.~\ref{tesigrafT6}.
We discuss the results for decreasing $a$.
For $a=0.1$, fig.\ref{tesigrafT6a}, the two methods give similar results 
except near zero temperature  where the critical capacity with   
optimization is slightly bigger than with self-control, like we expect
from the analysis at zero temperature. Decreasing the value of the pattern
activity to $a=0.01$, fig.\ref{tesigrafT6b}, self-control starts to work
less good for a bigger region of high temperatures but it is better at
lower temperatures. The curves in fig.~\ref{tesigrafT6d} show that at
high temperature the region of retrieval with self-control becomes
rather small when the activity is further lowered to $a=0.001$. We
also remark that 
for any choice of the pattern activity below $0.05$ there is a value of
the temperature where the two curves intersect. This is consistent with
the fact that at low activity in the limit of zero temperature
self-control works better than opimization for $a<0.05$. 
We conclude that, compared with optimization, self-control gives quite good 
results for activities in the range  $a\in [0.01, 0.05]$. When we want to
consider lower activities ($a=0.001$ and less) at arbitrary non-zero
temperature  self-control ceases to be a good method to control the
noise during the dynamics of the network.
In this case the temperature dependent externally chosen threshold
optimizing the mutual information function leads to better retrieval
qualities than the temperature independent self-control threshold.

         

\section{Concluding remarks}
In this paper we have studied the effects of a threshold in the gain
function on the parallel dynamics in layered neural networks with
variable activity. Such a threshold considerably enlarges the critical
capacity of the network. 
Two different types of thresholds are considered. The first one forces 
the neural
activity to be the same as the activity of the stored patterns at every
step of the retrieval process and adapts itself for this purpose in the
course of the time evolution. It provides a complete self-control mechanism.
The second optimizes the mutual information function in equilibrium. It
has to be given externally. 
For zero temperatures and low activity $a \leq 0.5$ it is found that
self-control
performs the best in considerably improving the storage capacity, the basin
of attraction and the mutual information content, exactly as for
extremely diluted models. And, in comparison with the optimization
method, it even gives a comparable improvement for higher activities.
Moreover, for non-zero temperatures self-control although being designed at 
temperature zero still gives quite good results for lower activities
$(a<0.5)$
that are bigger than $0.01$. Outside this region optimization done externally
for every loading and every temperature leads to better overall retrieval
qualities (except, obviously, near the critical capacity at zero
temperature). 
It is worth studying whether self-control can still be improved by
making it temperature dependent and/or whether optimization of the
mutual information content can be done in a self-controlled way.

\section*{Acknowledgments}
This work has been supported in part by the Research Fund of the
K.U.Leuven (Grant OT/94/9). The authors are indebted to
G.~Jongen for constructive discussions.
One of us (D.B.) thanks the Fund for Scientific
Research-Flanders (Belgium) for financial support.
    

\newpage

\end{document}